\documentclass[journal,onecolumn]{IEEEtran}
\ifCLASSINFOpdf
  \usepackage[pdftex]{graphicx}
\else
\fi
\usepackage{algorithmic}
\usepackage{algorithm}

\usepackage{hyperref}

\begin{document}
\title{TeleBTC: Trustless Wrapped Bitcoin}

\author{
    \IEEEauthorblockN{
        Mahyar Daneshpajooh, 
        Niusha Moshrefi, 
        Mahdi Darabi, 
        Sina Hashemi, 
        Mehrafarin Kazemi
    } 
    \\
    \IEEEauthorblockA{
        \{mahyar, niusha, mahdi, sinah,
        mehrafarin\}@teleportdao.xyz}
    \thanks{}% <-this % stops a space
}


\markboth{}%
{Shell \MakeLowercase{\textit{et al.}}: TeleBTC: Trustless Wrapped Bitcoin}



\maketitle

\begin{abstract}

This paper introduces TeleBTC, a fully decentralized protocol designed to wrap Bitcoin (BTC) on programmable blockchains. The creation of a decentralized wrapped BTC presents challenges due to the non-programmable nature of Bitcoin, making it difficult to custody BTCs in a decentralized way. Existing solutions have addressed this challenge by introducing an external layer of validators who take custody of users' BTCs. However, the security and decentralization of this layer are inferior to the underlying blockchains on which wrapped BTC is built. Moreover, the process of joining or leaving for a validator has become overly complex and expensive. To overcome these limitations, we propose a novel approach that eliminates the need for such an external layer by leveraging the light client bridge protocol. Additionally, we employ economic mechanisms such as incentivization and slashing, resulting in a secure and trust-minimized wrapped BTC solution. With TeleBTC, users can seamlessly transfer their BTC to other blockchains and utilize it within decentralized applications. Furthermore, they can unwrap their TeleBTC and reclaim the native BTC. To address the high costs associated with light client bridges, we present an optimistic approach that minimizes the cost. This approach significantly reduces the operational expenses of running the protocol.

\end{abstract}

% % Note that keywords are not normally used for peerreview papers.
% \begin{IEEEkeywords}
% IEEE, IEEEtran, journal, \LaTeX, paper, template.
% \end{IEEEkeywords}

% \fi
\IEEEpeerreviewmaketitle

\section{Introduction}
\IEEEPARstart{T}{he} advent of blockchain technology brought about the first blockchain-based system, Bitcoin \cite{Bitcoin}, in 2008 by Satoshi Nakamoto. Using the blockchain, Satoshi developed a peer-to-peer payment system allowing users to transfer money without relying on a centralized third party. Subsequently, Ethereum \cite{Ethereum} emerged, expanding upon the concepts introduced by Bitcoin and creating a decentralized computer capable of running arbitrary programs in a decentralized manner. These programs, referred to as smart contracts, operate without the need for a centralized server. Following Ethereum, numerous other programmable blockchains have emerged, each with its own focus, such as Decentralized Finance (DeFi) \cite{werner2022sok} and gaming, and improved features like scalability \cite{rocket2020scalable}, and speed.

Users of programmable blockchains have the opportunity to leverage decentralized applications (dApps) developed on these platforms, such as lending protocols and Decentralized Exchanges (DEX) \cite{10.1145/3570639}. However, Bitcoin users face a limitation as the Bitcoin blockchain lacks programmability, preventing direct access to these dApps. One potential solution for Bitcoin users is to exchange their BTC for alternative tokens on different blockchains to access the desired dApps. Nonetheless, this approach may not be suitable for many Bitcoin users who wish to maintain their exposure to the BTC price.

This challenge has prompted the community to explore solutions that enable the integration of Bitcoin with other programmable blockchains, allowing users to retain their BTC holdings while accessing the benefits of DeFi dApps and more. This has led to the emergence of blockchain bridges \cite{Zamyatin2019SoKCA}, which facilitate the interoperability \cite{interop} between different blockchains. Additionally, the concept of wrapping tokens has been introduced, enabling the representation of tokens from one blockchain to another.


 

\subsection{Blockchain Bridges}
To connect different blockchains, blockchain bridges have been developed to facilitate the transfer of tokens, NFTs, and data between them. These bridges come in different structures, offering varying levels of decentralization and security. There are two main categories of bridges: validator-based \cite{axelar} and light-client-based \cite{zkBridge}.

Validator-based bridges rely on external nodes to verify the data from one blockchain (called source chain) on another (called target chain). While this approach may be easier to establish, it sacrifices decentralization and security, particularly when the underlying blockchains are decentralized. On the other hand, light-client-based bridges leverage the smart contract to verify the proof of consensus of the source chain on the target chain. Therefore, they inherit the security and decentralization of the underlying blockchains, as they do not depend on external validators for data verification. This can incur high costs, especially when the target blockchain has high gas fees.

%\IEEEpubidadjcol

% \subsubsection{Subsubsection Heading Here}
% Subsubsection text here.

\subsection{Wrapped Tokens}

Locking the asset on one chain and minting an equivalent amount of it on another chain is called wrapping the asset \cite{wrappedAssets}. Wrapped assets are representations of original assets on other blockchains, enabling users to utilize them there. For instance, users can transfer their assets to a different blockchain in order to sell them on an exchange that provides more favorable rates. Another example is the option to deposit their assets into a protocol that offers higher yields compared to the protocols operating on the source blockchain. 

Users can redeem their native assets by burning the wrapped asset on the target chain. As long as each wrapped asset is fully backed by the native asset and users have the ability to redeem the native asset, the wrapped asset maintains an equivalent value to the native asset.

In a wrapped asset scheme, it is crucial to have a custodian responsible for locking the underlying asset until the corresponding wrapped tokens are burned. This ensures the security of the protocol. The custodian can be a centralized entity that is trusted not to move the assets. Alternatively, in a programmable blockchain, a smart contract can be utilized to lock the assets until specific conditions for unlocking are met. However, this poses a challenge in non-programmable blockchains like Bitcoin. In the case of Bitcoin, an address is needed to lock users' BTC. If this address is controlled by a single trusted entity, there is a risk that it could move the locked BTC and render the wrapped assets worthless.

One solution to address this problem is to create an address on Bitcoin that is controlled by multiple validators. In this approach, a consensus among a certain proportion of validators is required to sign a transaction and spend the locked assets. Therefore, a single validator or even a small number of colluding validators cannot move the users' locked BTC. However, achieving full decentralization in these solutions requires a significant number of validators, and their distribution should be genuinely decentralized. This can be challenging for projects to bootstrap and attract a large number of validators. With a small validator set, there is a risk of censorship where users' requests may be intentionally blocked, and collusion among validators could potentially lead to the theft of users' assets.

\subsection{Related Works}
WBTC \cite{wbtc}, the pioneering wrapped BTC, was created on the Ethereum blockchain to empower BTC holders to leverage their assets within the DeFi ecosystem of Ethereum. However, WBTC's highly centralized approach for minting and burning introduces risks and compromises decentralization principles. As a response to this, alternative approaches such as renBTC \cite{ren} and tBTC \cite{tbtc} have emerged, offering more decentralized solutions. RenBTC utilizes the RenVM network, which serves as a decentralized custodian for BTC. When wrapping BTC, users deposit their BTC to an address generated and controlled by the RenVM network. The RenVM network leverages Shamir's Secret Sharing \cite{shamir} to distribute key shares among its validators, ensuring the security of the deposited BTC. When users wish to redeem their BTC, RenVM validators collaborate to validate the request and facilitate the release of the BTC. Similarly, tBTC relies on external validators for security. These validators protect users' BTC through the use of threshold cryptography \cite{threshold}, and they rotate on a weekly basis to prevent collusion and fraudulent activities. In the event of compromised wallets, tBTC has an insurance backstop known as the coverage pool, which provides additional protection to users.

For wrapping BTC on other blockchains, the aforementioned solutions introduce an additional layer of validators. The security and decentralization of these protocols are confined to this specific layer. The validators lock some value that is susceptible to slashing in the case of misbehavior. The effectiveness of this layer's security relies on the economic value locked by the validators. In practice, the economic value within the validators' network is often low, resulting in significantly lower security compared to the security provided by the underlying blockchains. Consequently, many users are hesitant to compromise the security of their BTC by transferring it to other blockchains. This concern becomes particularly critical in the initial stages of a wrapped BTC protocol, as the security of the validator layer is extremely low. Moreover, these solutions face challenges when a validator decides to join or leave the protocol. Since validators hold a portion of the private key controlling the BTC, the BTCs must be transferred to a new address where the leaving validator has no access to the corresponding private key. This necessitates the movement of all BTCs to new addresses for every output associated with that validator, resulting in significant costs for the protocol. Similarly, when a new validator joins, they do not have access to the existing private keys, requiring the BTCs to be transferred to a new address associated with the joining validator.

\subsection{Contributions}
This paper presents TeleBTC, a fully decentralized wrapped BTC solution that empowers BTC holders to securely transfer their BTC to other blockchains and utilize it within various applications. To achieve this, we use the power of decentralized bridges to access the Bitcoin blockchain data on programmable blockchains. Then, we establish the wrapping solution on top of the bridge. Our research contributions include the following:

\begin{itemize}
  \item Designing a wrapped BTC protocol that eliminates the requirement for an external validator layer, ensuring that TeleBTC inherits the same level of security as the underlying blockchain.
  \item Introducing an optimized light-client bridge specifically tailored for the Bitcoin network. Our approach addresses the cost inefficiencies typically associated with light-client bridge solutions.
  \item Implementing the TeleBTC protocol on an EVM blockchain platform.
\end{itemize}

\section{TeleBTC Protocol}

\begin{figure}[!t]
\centering
\includegraphics[width=5in]{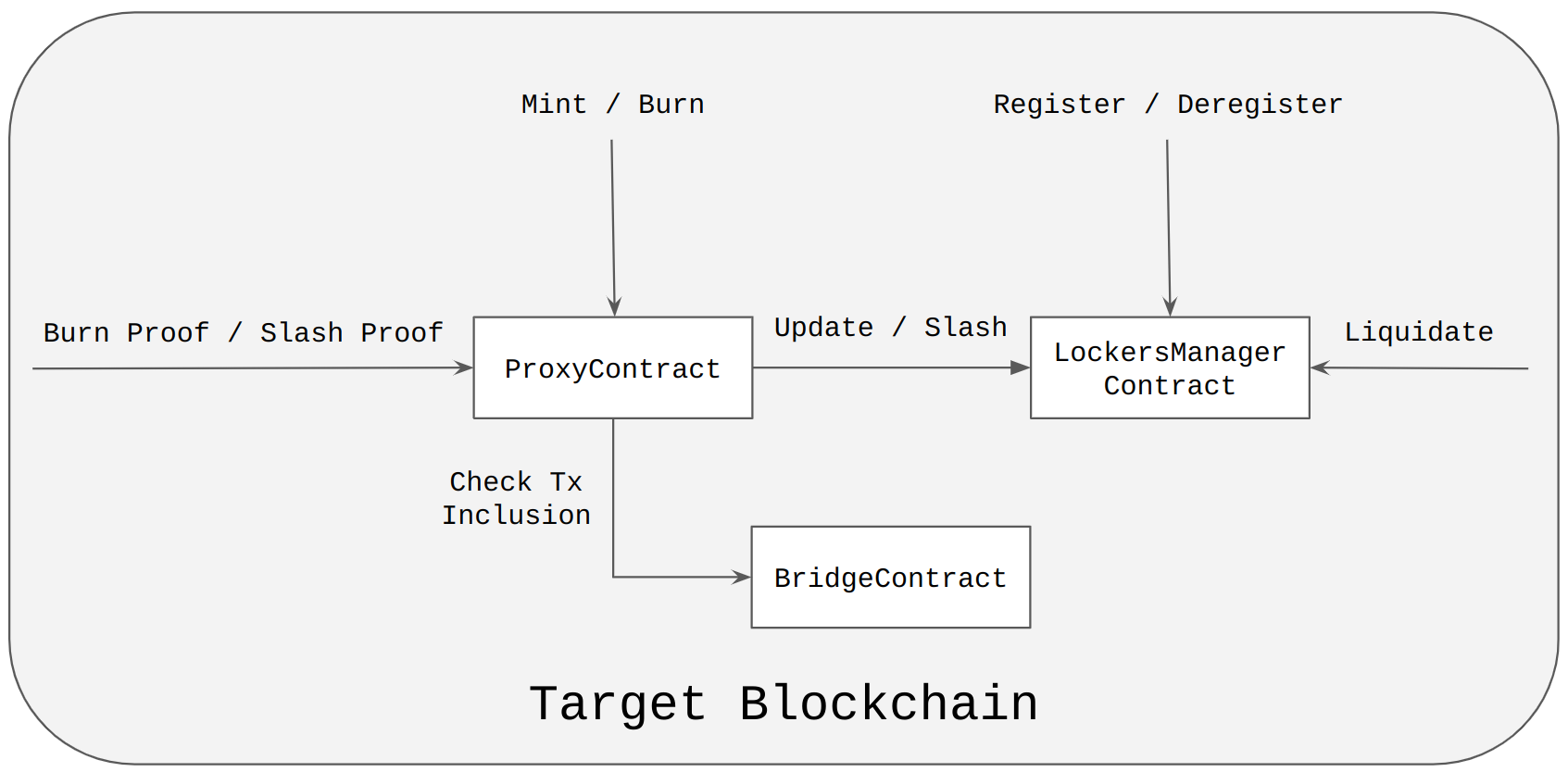}
\caption{Smart contract structure of the TeleBTC protocol. Mint and burn requests are processed by ProxyContract. The inclusion of transactions on Bitcoin is checked by BridgeContract. LockersManager contract handles the registration of Lockers and manages their collaterals.}
\label{contract-structure}
\end{figure}

In this section, we introduce TeleBTC, a decentralized protocol that empowers users to transfer their BTC seamlessly between the Bitcoin network and a programmable blockchain, referred to as the target chain. We also utilize the term TeleBTC to refer to the wrapped BTC within the context of the TeleBTC protocol. This protocol consists of four key elements: the light client bridge, Lockers, Teleporters, and Slashers. The smart contract structure of the protocol is shown in figure~\ref{contract-structure}.

The light client bridge plays a vital role in storing and validating Bitcoin data on the target chain, establishing a crucial link between the two networks. Within the protocol, Lockers are designated nodes responsible for securely holding users' BTC on the Bitcoin blockchain. To ensure their honest behavior, Lockers must lock collateral on LockersManagerContract, creating economic incentives for their reliability. In the event of misbehavior, Slashers submit proof of such misbehavior, resulting in slashing penalties for the malicious Locker. Teleporters are nodes that facilitate the seamless transfer of user requests from Bitcoin to the target chain. Their role involves collecting users' requests from Bitcoin, providing inclusion proof for them, and submitting them to TeleBtcContract on the target chain to mint TeleBTC, simplifying the overall process for users. Both Lockers and Teleporters receive fees for the services they provide. Importantly, these roles are open, meaning anyone has the opportunity to become a Locker or a Teleporter within the protocol.

\subsection{Light Client Bridge}
The light client bridge, known as BridgeContract, serves as an implementation of a Bitcoin light client on the target blockchain. Its primary objective is to enable the target blockchain to verify the inclusion of data sourced from Bitcoin. This is accomplished through the involvement of nodes known as Relayers. Relayers retrieve blocks from the Bitcoin blockchain and submit relevant data to BridgeContract for verification. BridgeContract verifies if the provided data introduces a state change that complies with the consensus mechanism of the Bitcoin blockchain or not. It also maintains the latest state of the Bitcoin canonical chain to determine which data has been finalized on the blockchain.

Light client bridges provide the highest level of security \cite{security} by ensuring that the submitted data is compatible with the consensus rules of Bitcoin. Additionally, they allow anyone to submit data for verification, enabling a decentralized process. As long as at least one honest Relayer submits data to the light client bridge, the bridge is synchronized with the Bitcoin blockchain.

The key functionality of BridgeContract is verifying whether a given transaction has been included in a specific block that has been finalized within the Bitcoin blockchain. A finalized block refers to a block that will permanently remain in the Bitcoin blockchain with an overwhelming probability.

There are several approaches to implementing a light client bridge for Bitcoin. One widely recognized protocol is the Simplified Payment Verification (SPV) protocol \cite{btcrelay}, where Relayers submit newly mined block headers of Bitcoin to BridgeContract. Another approach is the zkBridge \cite{zkrelay}, where Relayer nodes submit a batch of block headers along with a zero-knowledge proof that demonstrates the validity of the submitted block headers. By establishing a light client bridge, the protocol ensures a secure and decentralized mechanism for verifying Bitcoin transactions within the target blockchain. In section \ref{sectionIV} we explain how light client bridges work in more detail and propose solutions for optimizing their cost.

\subsection{Locker}

To wrap BTC, users are required to lock their BTC on the Bitcoin network. Conversely, to unwrap the wrapped BTC, users should be able to unlock the locked BTC. The security of the locking and unlocking mechanisms is critical to maintaining the price peg between the wrapped BTC and the native BTC. If users mint wrapped BTC without locking an equivalent amount of native BTC or unlock BTC without burning the corresponding amount of wrapped BTC, the price peg between these assets would be compromised.

A naive solution for asset locking is to rely on centralized custodians. However, this approach exposes users to the risk of asset theft that renders the wrapped asset valueless. Additionally, custodians may deny locking or unlocking assets for specific users, further compromising trust. Alternatively, users can send assets to a smart contract acting as a decentralized custodian. Unfortunately, due to the non-programmable nature of Bitcoin, locking BTCs using smart contracts is not feasible. To address these challenges, we introduce Lockers as decentralized custodians for users' BTCs. Each Locker functions as an independent node and is subject to slashing if they misbehave, which is the key distinction between Lockers and centralized custodians. Users lock their BTC by sending it to a designated Bitcoin address associated with a Locker. Subsequently, users can mint wrapped BTC on the target blockchain by providing proof of their BTC deposit to the Locker. To mitigate the risks of asset theft and user censorship, we implement monitoring of Lockers' activities. By incorporating economic guarantees, we ensure that Lockers behave honestly and maintain the security of users' assets.

On the target chain, we have a contract known as LockersManagerContract. To become a Locker within the system, a node must register with this contract. During the registration, Lockers provide their Bitcoin address, which users will utilize to lock their BTC by sending it to that address. Additionally, the node must lock collateral on the target chain. This collateral serves as a deterrent against malicious behavior and can be slashed if the Locker misbehaves. Lockers also can de-register from the protocol and retrieve their original collateral, provided they have not been slashed. The registration and de-registration of Lockers are managed by LockersManagerContract, making the protocol permissionless, allowing anyone to join or leave as a Locker.

\subsubsection{\textbf{Locker's Responsibilities}}
The main responsibilities of Lockers are twofold. Deviation from these responsibilities will result in slashing.

\paragraph{{\textbf{Safeguarding Locked BTCs}}}

To mint TeleBTC, users lock their BTCs by transferring them to Lockers. If a Locker transfers locked BTCs without receiving a valid request from a user to do so, a node (called Slasher) triggers ProxyContract. The Slasher provides ProxyContract with the Bitcoin transaction in which the fund transfer occurred. ProxyContract then calls BridgeContract to verify the inclusion of this transaction on Bitcoin and extracts relevant data from it. It checks whether the sender of the transaction was the Locker and retrieves the amount of transferred BTC. Then, it calls LockersManagerContract to slash the Locker's collateral. To disincentivize Lockers from stealing users' BTC, the amount of slashed collateral exceeds the amount of transferred BTC. Since the Locker can potentially steal the entire amount of locked BTC, the Locker’s collateral must be greater than the locked BTC at any time.

To fulfill this requirement, LockersManagerContract determines the $MintingCapacity$ of each Locker. $MintingCapacity$ represents the amount of TeleBTC that can be minted by that Locker while ensuring that the value of collateral remains greater than the value of locked BTC in its address. If the $MintingCapacity$ of a Locker becomes zero, the contract will not allow users to mint TeleBTC by utilizing this Locker. Users can still utilize other Lockers with sufficient $MintingCapacity$ to mint TeleBTC.

One challenge arises from price movements, as the value of collateral can decrease relative to the value of locked BTC. To address this, we implement a liquidation mechanism. During the process of liquidation, a Locker's collateral is sold for TeleBTC at a discounted rate, resulting in a healthy ratio between locked BTC and collateral. To prevent liquidation from occurring, Lockers can add extra collateral to the system at any time to maintain a healthy collateral position. Additionally, if the price moves in the opposite direction (i.e., If the price of the collateral increases in comparison to BTC), Lockers can withdraw part of their collateral from the system, still maintaining a healthy collateral position.

\paragraph{{\textbf{Processing Unlocking Requests}}}

To unlock their assets, users burn their TeleBTC on the target chain and request the Lockers to send them BTC on Bitcoin. Within the request, users specify their receiving address on the Bitcoin network. The assigned Locker responsible for the request will send the BTC to that address. However, there is a risk that the Locker may disregard the user's request, resulting in a potential loss of funds. In this case, the user has burnt some wrapped BTC on the target chain but has not received the native BTC on the Bitcoin network. 

To address this risk, a predetermined deadline is established after each unlocking request. The assigned Locker must send the assets on Bitcoin and provide proof of payment to ProxyContract within this timeframe. ProxyContract verifies the validity of the proof by utilizing BridgeContract. The transaction details, including sender, receiver, and transferred amount, are extracted and compared against the user's request. If the data aligns, the request is marked as processed. 

In the event that the Locker fails to provide the necessary proof before the deadline, the user can report the issue to the contract. Consequently, the contract initiates a slashing action against the Locker's collateral, transferring it to the user. Importantly, the amount of collateral slashed exceeds the value of the burnt BTC, disincentivizing Lockers to not fulfill users' unlock requests. It is important to consider that the deadline for executing requests should strike a balance. It should be long enough to allow Lockers to provide payment proof without being hindered by blockchain network congestion. However, it should not be excessively long to ensure that users receive their assets within a reasonable timeframe.

During the unlock process, Lockers are required to transfer a portion of the locked BTC to a user. This transfer should not be perceived as a malicious transfer by the contract. To this end, LockersManagerContract stores the ID of transactions that are provided by Lockers as payment proof. If someone later submits such a transaction as malicious, the contract will discard it. However, after sending the payment and before submitting the payment proof, the Locker might be slashed for deviating from safeguarding BTCs. To prevent this, Slashers cannot report the payment transaction as malicious in the predetermined timeframe when the Locker is still creating the proof for the payment. The contract strictly enforces that slashing can only occur for transactions that are older than a predetermined time, ensuring that the Locker has sufficient time to provide the necessary proof and avoid any unjust slashing.

\subsubsection{\textbf{Lockers Fee}}

To incentivize nodes to become Lockers in the protocol, a percentage of the total minted and burnt BTC is allocated to the corresponding Locker. However, one challenge associated with Lockers is that if the demand for minting significantly outweighs the demand for burning, the Locker capacity may be reached, potentially causing the system to halt minting. To address this issue, Lockers can increase their collateral or new Lockers can join the system to increase $MintingCapacity$. 

To effectively handle such situations, we employ a dynamic adjustment mechanism for $MintingFee$ and $BurningFee$, which is based on the ratio between the total $MintingCapacity$ and the total locked BTC. When $MintingCapacity / locked BTC$ approaches zero, indicating that a significant portion of the capacity is being utilized, $MintingFee$ increases. This increase serves as an incentive for more Lockers to join the system and increase the system capacity.

Conversely, $BurningFee$ is proportional to this ratio. As the $MintingCapacity$ decreases, $BurningFee$ decreases as well. This adjustment makes it more cost-effective for users to burn their TeleBTC to unlock native BTC when $MintingCapacity$ is low. By dynamically adjusting $MintingFee$ and $BurningFee$, we aim to maintain the balance between minting and burning demand, ensuring the smooth operation of the system.

\begin{figure}[!t]
\centering
\includegraphics[width=5in]{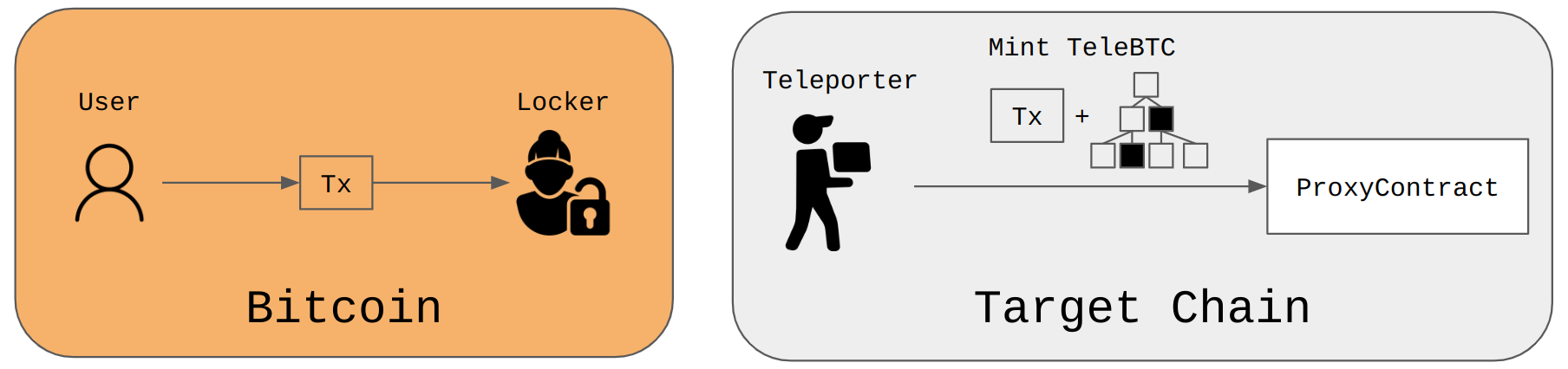}
\caption{To wrap BTC, the user transfers them to a Locker. Once this transaction is confirmed on the Bitcoin network, a Teleporter collects it and submits the transaction along with the inclusion proof to ProxyContract.}
\label{mintFig}
\end{figure}

%%% Wrapping BTC %%%
\begin{algorithm}\floatname{algorithm}{Procedure}
\caption{\textsc{Wrapping BTC}}
\label{wrapBtc}
\begin{algorithmic}[1]

\STATE X $\leftarrow$ Amount of BTC to wrap
\STATE The user obtains the list of Lockers, including their Bitcoin addresses and $MintingCapacity$, from LockersManagerContract.
\STATE The user sends $X$ BTC to the Locker with the highest $MintingCapacity$ exceeding $X$, specifying the receiver address and the percentage fee designated for the Teleporter.
\STATE Once the transaction is confirmed on Bitcoin, a Teleporter generates proof for it and submits both the proof and the transaction to ProxyContract.
\IF{The transaction ID has not been recorded before}
\STATE Records the transaction ID.
\STATE Verifies the proof using BridgeContract.
\STATE Extracts the Locker address, sent amount, receiver address, and TeleporterFee from the transaction.
\IF{The Locker address is valid AND the Locker's $MintingCapacity$ exceeds $X$}
\STATE Mints $X$ wrapped BTC and updates LockersManagerContract.
\STATE Transfers $TeleporterFee * X$ to the Teleporter, $MintingFee * X$ to the Locker, and the remaining amount to the specified receiver address.
\ENDIF
\ENDIF

\end{algorithmic}
\end{algorithm}

\subsection{Teleporter}
When a user requests to mint TeleBTC on the target chain, the request must be transmitted to ProxyContract. The user sends a transaction to move BTC to a Locker's address to lock it there, then needs to trigger the target chain to mint the equivalent amount of TeleBTC. However, due to the approximately one-hour finalization time on the Bitcoin network, users cannot submit the request to this contract immediately after sending the Bitcoin transaction. To address this issue and avoid requiring users to manually submit the request after an hour, we have introduced Teleporters. Teleporters act as intermediaries that collect users' requests from Bitcoin and submit them to ProxyContract on the target chain. They also eliminate the need for users to interact directly with two different chains, which would require holding the native coin of each blockchain to cover transaction fees. Instead, users only need to interact with the Bitcoin network for minting TeleBTC. 

To execute a request, a Teleporter generates a Merkle inclusion proof \cite{merkle} that demonstrates the inclusion of the request (transaction) in a finalized block in the Bitcoin network. Teleporter submits this proof to ProxyContract on the target chain. ProxyContract records the transaction ID of the submitted request to prevent multiple submissions of the same request. Within the request sent on Bitcoin, users specify the receiver address and the percentage of tokens that will serve as the fee for the Teleporter. This fee incentivizes Teleporters to submit the request on behalf of the user. The fee amount covers the cost of executing the request on the target chain plus an extra reward for the Teleporter. It should be noted that users have the ability to submit their own requests to ProxyContract, even if no Teleporter submits the request due to factors like low fees.

\subsection{Slasher}

Slashers play a critical role in ensuring the accountability of Lockers within the network. Anyone has the potential to become a Slasher. Their main task is to slash Lockers' collateral if they engage in misconduct, backed by providing proof of the misbehavior. The slashing mechanism is a vital component that guarantees the safety and liveness of the protocol.

Slashers can slash Lockers in two scenarios. Firstly, if Lockers deviate from their responsibility of safeguarding BTC, and secondly, if Lockers fail to process unlocking requests promptly. In either case, the Slasher reports the misconduct to ProxyContract, which verifies the slashing condition and initiates the slashing process against the deviating Locker. As a reward for their contribution, the contract allocates a portion of the slashed collateral to the Slasher, providing them with the necessary incentive to continue fulfilling their role.

In the subsequent section, we will delve into the details of the slashing mechanisms, explaining how they ensure the overall safety and liveness of the protocol.

\subsection{All Together}

\begin{figure}[!t]
\centering
\includegraphics[width=5in]{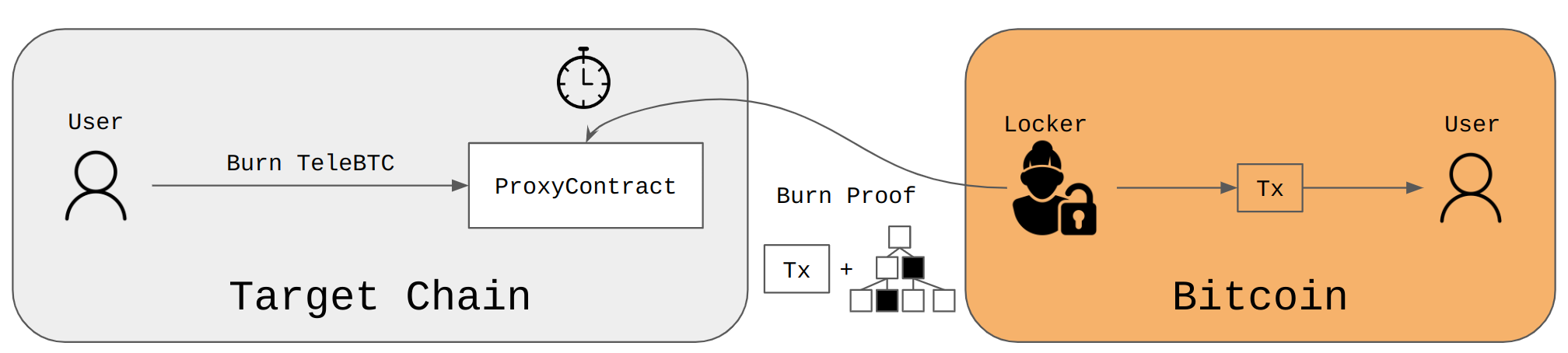}
\caption{The process of unwrapping TeleBTC involves the user sending a burn request and the TeleBTC token to ProxyContract. Once this request is received, a timer starts for the Locker. The Locker is responsible for sending the corresponding BTC to the user on the Bitcoin network and providing proof of payment to ProxyContract before the deadline.}
\label{burnFig}
\end{figure}

Now, let's bring all the components together to explain the functioning of the TeleBTC protocol. The protocol encompasses two primary functionalities: wrapping (Figure~\ref{mintFig}) and unwrapping (Figure~\ref{burnFig}). 

To wrap BTC into TeleBTC tokens, the user initiates the process by sending a transaction on the Bitcoin blockchain (Procedure~\ref{wrapBtc}). This transaction moves BTC to a designated Locker address and includes the user's specified address on the target chain where TeleBTC will be sent. Once the transaction is confirmed, a Teleporter collects it and submits the transaction along with proof of its inclusion on Bitcoin, to ProxyContract on the target chain. ProxyContract verifies the request data and ensures its inclusion and finalization on Bitcoin by querying BridgeContract. Upon successful verification, TeleBTC tokens are minted, and the Teleporter and Locker receive their respective rewards. Finally, the remaining TeleBTC tokens are transferred to the user's specified address on the target chain, completing the wrapping process.

To initiate the process of unwrapping TeleBTC back into BTC, the user submits an unlock request to ProxyContract (Procedure~\ref{unwrapTelebtc}). Within the request, the user includes the TeleBTC tokens, specifies the recipient address on the Bitcoin blockchain, and designates a Locker to handle the request. Upon receiving the request, ProxyContract burns the TeleBTC tokens and records the unwrap request. Subsequently, a timer is set for the Locker, providing a deadline for processing the request. The Locker then proceeds to transfer the corresponding BTC to the user's recipient address on Bitcoin. Before the deadline expires, the Locker submits payment proof to ProxyContract. ProxyContract verifies the validity of proof using BridgeContract and extracts the relevant details to ensure the request has been accurately processed.

Finally, to ensure the safety and liveness of the protocol, liquidation and slashing mechanisms are established for Lockers. Slasher nodes take care of slashing the participants in case of any misbehavior. These mechanisms are explained and analyzed in the next section.

The TeleBTC protocol is implemented and tested between an EVM-compatible blockchain and Bitcoin. The code can be found \href{https://github.com/TeleportDAO/teleswap-contracts}{here}.

%%% Unwrapping TeleBTC %%%
\begin{algorithm}\floatname{algorithm}{Procedure}
\caption{\textsc{Unwrapping TeleBTC}}
\label{unwrapTelebtc}
\begin{algorithmic}[1]

\STATE X $\leftarrow$ Amount of TeleBTC to unwrap
\STATE The user obtains the list of Lockers with their Locked BTC amount from LockersManagerContract.
\STATE The user finds the address of the Locker with the highest locked BTC amount that exceeds X.
\STATE The user requests ProxyContract to burn $X$ TeleBTC by providing the Locker address and the receiver address.
\STATE ProxyContract burns $Y = X * (1 - BurningFee)$ TeleBTC 
 and sends $X - Y$ to the Locker. It also updates LockersManagerContract.
\STATE ProxyContract starts a timer for the Locker. 
\STATE The Locker sends $Y$ BTC to the receiver address, waits for confirmation and then provides the payment proof to ProxyContract.
\STATE ProxyContract verifies the proof using BirdgeContract and extracts the sender, receiver, and amount from the transaction.
\STATE If the sender of the transaction is identified as the Locker, the receiver is the receiver address, the transferred amount is Y, and the deadline for execution has not passed, the contract marks the request as processed and records the payment transaction ID.

\end{algorithmic}
\end{algorithm}

\section{Protocol Analysis}

In this section, we analyze the mechanisms that ensure the safety and liveness properties of the TeleBTC protocol. We begin by providing definitions for safety and liveness in the context of wrapped tokens. Subsequently, we outline how we achieve these properties through the implementation of economic guarantees.

\subsection{Safety}

The safety of TeleBTC as a wrapped asset depends on its backed asset. We assert that TeleBTC is safe as long as each TeleBTC is backed by at least one native BTC. To uphold this property, several cases should be considered: no one should be able to mint 1 TeleBTC by locking less than 1 BTC, as doing so would violate the safety requirement. Additionally, no one should be able to obtain 1 BTC by burning less than 1 TeleBTC. As long as the number of minted TeleBTC does not exceed the number of locked BTC, the protocol remains safe.

In our protocol, a user who sends 1 BTC to a Locker can provide the transaction to ProxyContract to mint 1 TeleBTC. This transaction cannot be used twice, as the unique transaction ID is recorded in ProxyContract, preventing double minting. Furthermore, the user cannot mint TeleBTC without first sending BTC to a Locker. If the transaction provided by the user hasn’t been finalized on the Bitcoin network or if the BTC wasn’t transferred to a Locker (it was transferred to a non-Locker address), ProxyContract will detect these inconsistencies and discard the minting request.

To ensure safety, we must ensure that Lockers have no incentive to steal locked BTCs. This means that whenever a Locker transfers 1 BTC to a user, the user must have burned at least 1 TeleBTC beforehand. To enforce this, Lockers are mandated to lock collateral within the protocol, subjecting them to potential slashing if the condition is violated. At any given time, the value of collateral for a Locker must exceed the value of the locked BTC. Failure to meet this requirement would create an incentive for the Locker to steal locked BTCs. To ensure this property, we implement a liquidation mechanism that addresses fluctuations in the price of collateral relative to BTC.

\subsubsection{\textbf{Liquidation}}

The liquidation mechanism within the protocol ensures that the value of collateral always exceeds the value of locked BTC for each Locker. Assuming the value of the collateral is represented by $CollateralValue$, the maximum amount of locked BTC ($MaxLockedBtcValue$) that can be held by the Locker is determined by $CollateralValue$ over  $CollateralizationRatio$, where $CollateralizationRatio$ is a number greater than 1. Due to price fluctuations, $LockedBtcValue$ for a Locker could surpass $MaxLockedBtcValue$. If price fluctuations are significant, $LockedBtcValue$ could even exceed $CollateralValue$. To prevent this, the protocol incorporates a liquidation mechanism that gets triggered when $LockedBtcValue$ exceeds $CollateralValue$ over $LiquidationRatio$. $LiquidationRatio$ is a number between 1 and the $CollateralizationRatio$, allowing users to liquidate the Locker before the $CollateralValue$ becomes lower than $LockedBtcValue$. 

Users can initiate the liquidation process by sending TeleBTC to the contract in exchange for the Locker's collateral (Algorithm~\ref{liquidation}). This process reduces the Locker's collateral, but overall improves the health of the system by burning the collected TeleBTC. The contract incentivizes user participation in liquidation by offering collateral at a discounted rate of $DiscountRatio$. The liquidation remains active until the $CollateralValue / LockedBtcValue$ becomes greater than $CollateralizationRatio$ again. After this point, the contract disallows further liquidation of the Locker to avoid inflicting additional losses. It is important to mention that since $CollateralizationRatio$ is greater than $LiquidationRatio$, this prevents immediate recurring liquidation.

Assuming the user purchases a value of the Locker's collateral with TeleBTC, denoted as $BoughtValue$, we aim to demonstrate that this action contributes to a healthier system. This is reflected in an increased ratio of $CollateralValue$ to $LockedBtcValue$. Following the liquidation, $CollateralValue$ decreases by $BoughtValue$, while the $LockedBtcValue$ decreases by $BoughtValue$ multiplied by $DiscountRatio$, as the user acquires the collateral at a discount. Thus, we need to establish the condition:

\begin{equation}
\frac{CollateralValue - BoughtValue }{LockedBtcValue - BoughtValue * DiscountRatio }> \frac{CollateralValue}{LockedBtcValue}
\end{equation}

This equation simplifies to: $DiscountRatio > LockedBtcValue / CollateralValue$. Since we assume that liquidation occurs before $CollateralValue$ becomes lower than $LockedBtcValue$, the right-hand side of the equation is less than 1. Therefore, any $DiscountRatio$ within this range promotes a healthier system. Additionally, considering that liquidation takes place after the $CollateralValue / LockedBtcValue$ falls below $LiquidationRatio$, $DiscountRatio$ should be greater than $1 / LiquidationRatio$. Ideally, $DiscountRatio$ should be close to 1 to ensure the system remains robust during high fluctuations. However, a $DiscountRatio$ closer to 1 also means there is less incentive for users to participate in the liquidation process.

%%% Liquidation %%%
\begin{algorithm}\floatname{algorithm}{Procedure}
\caption{\textsc{Liquidation}}
\label{liquidation}
\begin{algorithmic}[1]

\STATE The user initiates the liquidation of a Locker by sending a value of $X$ TeleBTC to LockersManagerContract.
\IF{$\frac{CollateralValue}{LockedBtcValue} < LiquidationRatio$ AND $\frac{CollateralValue - X / DiscountRatio}{LockedBtcValue - X} < CollateralizationRatio$}
\STATE LockersManagerContract burns TeleBTC tokens and transfers $X / DiscountRatio$ worth of collateral to the user.
\ELSE 
\STATE The user's request is discarded.
\ENDIF

\end{algorithmic}
\end{algorithm}

\subsubsection{\textbf{Slashing for Safety}}

By utilizing liquidation, we ensure that at any given moment, the value of the collateral is greater than the locked BTC, thereby holding the Locker accountable for the locked BTCs. If the Locker steals some BTC, leading to a lower number of locked BTC than minted TeleBTCs, we need to slash the Locker's collateral. In this scenario, we slash an amount of the Locker's collateral greater than the stolen value. Since the Locker's collateral value is greater than the locked BTC, this is feasible. In the event of slashing, the Slasher sends a corresponding number of TeleBTCs to the contract, equivalent to the number of stolen BTCs, and in return receives the Locker's collateral with a discount. The sent TeleBTC is then burned by the contract. Essentially, by burning the same amount of TeleBTC that was stolen by the Locker, we restore the balance between the number of locked BTCs and minted TeleBTCs. 

In summary, through the mechanisms of slashing and liquidation, we ensure that the Lockers are accountable for the locked BTCs and prevent them from stealing BTCs.

\subsection{Liveness}

Another crucial requirement for a wrapped asset is liveness, ensuring that users can process their wrap and unwrap requests eventually. This guarantees that users do not lose their assets; if they lock their BTC, they will receive the corresponding TeleBTC, and if they burn their TeleBTC, they will receive the corresponding BTC.

\subsubsection{\textbf{Capacity Reservation}}

During the wrapping process, users send their BTC to a Locker, and the corresponding transaction along with the inclusion proof is provided to ProxyContract. Users do not require any permission to mint TeleBTC; they simply need to provide valid payment proof to ProxyContract. Assuming that BridgeContract maintains liveness, meaning that the Bitcoin block headers get finalized on it eventually, users can provide valid inclusion proof. The only factor that can prevent the minting of TeleBTC is the MintingCapacity of the Locker. If there are requests before the user's request that already utilized the MintingCapacity, and there is not enough capacity left for the user, the user cannot mint TeleBTC. The user must wait for some capacity to be freed up (e.g. some users burn their TeleBTC) or for the Locker to add more collateral. Once there is unused capacity, the user can resubmit the transaction to ProxyContract to mint TeleBTC. 

To address the concern raised, one effective approach is to implement capacity reservations. When a user intends to mint TeleBTC, they can notify ProxyContract in advance before locking their BTC. ProxyContract will accept the reservation if sufficient capacity is available. Subsequently, the user is provided with a predefined time window during which they can submit their minting requests and utilize the reserved capacity. This mechanism ensures that the locked BTC will be minted within a reasonable timeframe, thereby mitigating potential liveness issues.

To prevent users from reserving capacity without the intention to use it, an additional measure can be implemented. Users can be required to lock a small collateral as a commitment. If they fail to mint TeleBTC before their reservation's deadline, their collateral will be subject to slashing, and their reservation will be canceled. This discourages users from making reservations they do not intend to fulfill, promoting more efficient and fair utilization of the capacity.

\subsubsection{\textbf{Slashing for Liveness}}

When users send requests to unwrap their TeleBTC, the Locker should send them BTC on the Bitcoin network. However, there is a possibility that the Locker may ignore a request and not send the BTC. To mitigate this risk, we utilize a slashing mechanism for liveness.

After the user's unwrap request is submitted on the smart contract, the Locker has a predefined deadline to send the unwrapped BTC to the user and submit its proof to ProxyContract. The proof shows that the Locker has transferred the needed BTC to the address determined by the user. If the Locker fails to provide valid proof before the deadline, the user can invoke the contract to slash the Locker. In this case, the contract sends $BurntValue / DiscountRatio$ from the Locker's collateral to the user. This mechanism discourages Lockers from neglecting users' burn requests. Since the amount of Locker collateral is greater than the total locked BTC, the Locker is encouraged to process unwrap requests, otherwise, it will face being slashed for a greater amount.

In summary, by incorporating capacity reservations, the wrapping process ensures liveness. Users can reserve capacity in advance, allowing for efficient minting of TeleBTC. Moreover, the time-bound unwrap process, along with the slashing mechanism, ensures the timely transfer of unwrapped BTC to users while incentivizing Lockers to fulfill their responsibilities. These measures collectively contribute to the overall reliability of the TeleBTC protocol.

\section{Optimization}\label{sectionIV}

One of the key components of the TeleBTC protocol is the light client bridge. This bridge plays a crucial role in verifying whether a specific transaction is included in the Bitcoin blockchain. The bridge is maintained by Relayer nodes that synchronize it with the Bitcoin blockchain. However, it's important to note that maintaining this bridge incurs a cost, as both verifying and storing data on blockchains can be expensive. In this section, we will discuss methods to make the bridge more cost-efficient, thereby making the overall operation of TeleBTC more affordable. We will begin by explaining the SPV (Simplified Payment Verification) bridge protocol, which is a well-known light client protocol for Bitcoin. Subsequently, we will explore ways to optimize this protocol to further reduce costs without compromising security and decentralization.

\subsection{SPV Bridge}
In the SPV bridge protocol, Relayers obtain block headers from the Bitcoin network and submit them to BridgeContract on the target chain. BridgeContract is responsible for verifying and finalizing these block headers. The verification process includes several checks to ensure the validity of each block header:

\begin{itemize}
  \item Previous Submission Check: BridgeContract verifies that the block header has not been previously submitted, preventing duplicate entries.
  \item Parent Hash Verification: The block header is validated to ensure it correctly references the previous block header through the parent hash field.
  \item Target Difficulty Validation: If the difficulty has not changed, BridgeContract confirms that the target difficulty of the block header matches that of the previous block. In the case of a new difficulty epoch, the target difficulty gets appropriately updated.
  \item Proof of Work (PoW) Check: The block header's work is checked by ensuring that the double hash of the block header is less than the target difficulty. This confirms that the block satisfies the required computational effort.
\end{itemize}

These checks are also performed by Bitcoin full nodes when storing a new block in their storage. Once a block header passes these checks, it is considered a valid block header by BridgeContract, which then stores it. However, at this stage, the block is not considered finalized and may still get slipped out of the canonical chain of Bitcoin. For a block header to get finalized, a specific number of consecutive block headers, known as the FinalizationNumber, need to be included in the chain on top of it. This requirement significantly reduces the chances of the block being disregarded. This concept aligns with how Bitcoin nodes consider a block as finalized. BridgeContract continually checks for finalized blocks after each new block is added. If the newly added block with the height $BlockHeight$ has a higher height than the previously submitted blocks, BridgeContract traces back to find the ancestor block at height $BlockHeight - FinalizationNumber$. It then removes all other blocks at that height from the blockchain, retaining only the ancestor block. Users can now utilize this finalized block with confidence that it will not get reverted in the Bitcoin canonical chain.

\subsection{Optimistic Bridge}

In the SPV bridge, Relayers are responsible for submitting each newly mined block header of Bitcoin to BridgeContract, where they undergo validation. This process incurs costs related to both block header submission and verification.

\subsubsection{\textbf{Reducing Verification Cost}}

To reduce the verification cost, we propose an optimistic approach. Instead of verifying each submitted block header, verification is only performed when a disputer challenges a block header. Disputers can challenge a header in $ChallengePeriod$, which gets started right after submitting the header. If no challenges are made within this period, the block is marked as valid, and its validity cannot be disputed in the future. In the event of a challenge, $ProofPeriod$ gets started. During this period, the Relayer who submitted the header must provide proof of its validity. If valid proof is provided within the period, the block header is marked as valid. Otherwise, the header is marked as invalid. 

To ensure incentive compatibility, Relayers are required to deposit collateral for each block header they submit. If a submitted block header is marked as valid, the collateral is returned to the Relayer. However, if the block is marked as invalid, the collateral gets slashed. Disputers also need to put up collateral before challenging a block header. If the block is determined to be invalid, they receive their collateral back along with the collateral of the Relayer as a reward. Otherwise, they lose their collateral to the Relayer. This approach compensates for the honest player's proof/challenge submission cost and provides them with additional rewards.

When a header is challenged, BridgeContract verifies several conditions, including the validity of the connection between headers, matching target difficulty with the previous header, and sufficient PoW. These checks are similar to those performed in the SPV bridge. 
By implementing this optimization on top of the SPV bridge, we maintain the same level of security while introducing an additional assumption: the presence of at least one honest disputer within the network.

One challenge of the optimistic approach is the introduction of delay. In the SPV bridge, a block header can be submitted as soon as it is mined on Bitcoin. However, in the optimistic bridge, a block header can only be submitted on top of another valid header, and header verification itself incurs a delay. In the normal case, a block header is verified after $ChallengePeriod$. In the case of challenges, verification occurs in $ProofPeriod$. To minimize the impact of delay, appropriate durations must be chosen for these periods. The average time between blocks in Bitcoin is 10 minutes. There exists a $ChallengePeriod$ smaller than this value, which, on average, does not cause any extra delay for the system. It should be noted that $ChallengePeriod$ cannot be too short, as that may prevent a disputer from challenging an invalid block if the blockchain is overloaded, resulting in the verification of the invalid header.

\subsubsection{\textbf{Reducing Submission Cost}}

Another cost associated with the SPV bridge is the storage of the entire block header data. However, BridgeContract only needs the Merkle roots of transactions (which is part of the block header structure) to check transaction inclusion proofs. So, to further reduce the bridge's cost, we propose storing only the Merkle roots in BridgeContract. 

Instead of submitting the entire block header, the Relayer would only need to submit the Merkle roots to the contract (Algorithm~\ref{submit-root}). Each newly submitted Merkle root should be on top of a previously verified root. In the event that a submitted Merkle root is challenged (Algorithm~\ref{challenge-root}), the Relayer would be required to provide the complete block header for both the challenged root and the verified root.

To prevent Relayers from submitting valid block headers with a high-difficulty target (which is computationally easier to construct), the difficulty target for each difficulty epoch is stored. This helps the contract to verify that the block headers provided by the Relayer have sufficient PoW. In order to store the difficulty target of epochs, Relayers must provide the timestamps of the first and last blocks of each difficulty epoch. This information allows BridgeContract to calculate the new difficulty target for the subsequent epoch based on the provided timestamps.

One potential attack is mining the last block of an epoch with an abnormally large timestamp. This would result in a very low difficulty for the next epoch, making it easier for an attacker to submit valid block headers matching that difficulty. To mitigate this, the contract will reject headers with timestamps much higher than that of the target blockchain. This prevents Relayers from manipulating the difficulty target.

%%% Submitting %%%
\begin{algorithm}\floatname{algorithm}{Procedure}
\caption{\textsc{Submitting a Merkle Root}}
\label{submit-root}
\begin{algorithmic}[1]

\STATE The Relayer retrieves $LastSubmittedHeight$ from BridgeContract and queries all the submitted Merkle roots at that height.
\STATE The Relayer obtains the Merkle root of the $LastSubmittedHeight$ block header from the Bitcoin blockchain.
\IF {The Merkle root exists in the submitted Merkle roots}
\IF{It is either verified or ChallengePeriod has passed}
\STATE The Relayer acquires the Merkle root of $LastSubmittedHeight + 1$ from the Bitcoin blockchain and submits it to BridgeContract along with the required collateral. 
\IF {The height of the Merkle root is at $2016k$ or $2016k + 1$}
\STATE The Relayer also submits the corresponding timestamp of the block header.
\ENDIF
\ELSE
\STATE The Relayer waits until $ChallengePeriod$ has passed or the Merkle root becomes verified.
\ENDIF
\ELSE 
\STATE {The Merkle root is deemed invalid, and the collateral provided by the Relayer, along with the original collateral, is transferred to the disputer.}
\ENDIF

\end{algorithmic}
\end{algorithm}

%%% Challenging %%%
\begin{algorithm}\floatname{algorithm}{Procedure}
\caption{\textsc{Challenging a Merkle Root}}
\label{challenge-root}
\begin{algorithmic}[1]

\STATE The disputer initiates a challenge by providing the required collateral to BridgeContract. If $ChallengePeriod$ is still active and the Merkle root in question has not been verified, BridgeContract begins $ProofPeriod$.
\STATE The Relayer obtains the block header corresponding to the challenged Merkle root, as well as the block header from the previous height, and submits them to BridgeContract.
\IF {The block headers are submitted within $ProofPeriod$ AND The Merkle roots match the stored Merkle roots in BridgeContract AND The block headers are correctly linked AND The target difficulty matches the required difficulty AND They meet the PoW requirements.}
\STATE BridgeContract marks the Merkle root as verified. 
\STATE The collateral provided by the disputer, along with the original collateral, is then transferred to the  Relayer.
\ELSE 
\STATE The Merkle root is deemed invalid.
\STATE The collateral provided by the Relayer, along with the original collateral, is transferred to the disputer.
\ENDIF

\end{algorithmic}
\end{algorithm}

\section{Conclusion}

In this paper, we introduced TeleBTC, a trustless protocol designed for wrapping and unwrapping BTC. TeleBTC enables users to transfer their BTC to programmable blockchains and utilize them in decentralized applications. Existing wrapped BTC solutions rely on external networks of validators to safeguard users' BTC holdings and facilitate cross-chain transactions. However, these additional layers often exhibit lower security and decentralization compared to the underlying blockchains. In contrast, TeleBTC achieves a high level of security and decentralization by eliminating the need for such intermediary layers.

TeleBTC achieves this by leveraging a light client bridge, which enables decentralized access to Bitcoin data on another chain. Furthermore, the TeleBTC protocol incorporates economic mechanisms designed to ensure the safety and liveness of the protocol. This reassures BTC holders that they can transfer their assets to other chains without risking their security or compromising decentralization. Additionally, we propose an optimization approach aimed at reducing the costs associated with the light client bridge. This approach only requires the submission of the Merkle root of transactions instead of the entire block header data. It also incorporates an optimistic approach for data verification that reduces the overall cost of the bridge.

Overall, TeleBTC presents a promising solution for users seeking to utilize their BTC in decentralized applications while maintaining the security and decentralization properties of the underlying blockchain. The protocol's elimination of additional layers and its lightweight client bridge demonstrates its potential to address the limitations of existing wrapped BTC solutions. Further research and development in this area will undoubtedly contribute to the advancement of trustless cross-chain protocols and the broader adoption of decentralized applications.

% or


% \appendices

% % you can choose not to have a title for an appendix
% % if you want by leaving the argument blank
% \section{}

% % use section* for acknowledgment
% \section*{Acknowledgment}


\ifCLASSOPTIONcaptionsoff
  \newpage
\fi



\bibliographystyle{IEEEtran}
\bibliography{Reference}
\end{document}